# Domain configurations in Co/Pd and $L1_0$-FePt nanowire arrays with perpendicular magnetic anisotropy

Pin Ho,[1*] Kun-Hua Tu,[1] Jinshuo Zhang,[1] Congli Sun,[2] Jingsheng Chen,[3] George Liontos,[4], Konstantinos Ntetsikas,[4] A. Avgeropoulos,[4] Paul M. Voyles,[2] Caroline A. Ross[1]

[1]Department of Materials Science and Engineering, Massachusetts Institute of Technology, MA, USA
[2]Department of Materials Science and Engineering, University of Wisconsin, Madison, WI, USA
[3]Department of Materials Science and Engineering, National University of Singapore, Singapore, Singapore
[4]Department of Materials Science and Engineering, University of Ioannina, University Campus – Dourouti, Ioannina, Greece

**Abstract:**

Perpendicular magnetic anisotropy [Co/Pd]$_{15}$ and $L1_0$-FePt nanowire arrays of period 63 nm with linewidths 38 nm and 27 nm and film thickness 27 nm and 20 nm respectively were fabricated using a self-assembled PS-b-PDMS diblock copolymer film as a lithographic mask. The wires are predicted to support Neel walls in the Co/Pd and Bloch walls in the FePt. Magnetostatic interactions from nearest neighbor nanowires promote a ground state configuration consisting of alternating up and down magnetization in adjacent wires. This was observed over ~75% of the Co/Pd wires after ac-demagnetization but was less prevalent in the FePt because the ratio of interaction field to switching field was much smaller. Interactions also led to correlations in the domain wall positions in adjacent Co/Pd nanowires. The reversal process was characterized by nucleation of reverse domains, followed at higher fields by propagation of the domains along the nanowires. These narrow wires provide model system for exploring domain wall structure and dynamics in perpendicular anisotropy systems.

*Corresponding author: P. Ho, E-mail: hopin@mit.edu

Self-assembly provides a set of pathways for the synthesis of functional nanostructures such as magnetic nanowires or nanodots. In particular the self-assembly of block copolymer (BCP) thin films generates periodic arrays of microdomains which can serve as a template for the patterning of nanoscale structures that may be useful in device applications.[1–7] Compared to optical lithography with its limited resolution and electron beam lithography with its limited throughput, BCP lithography offers advantages in the fabrication of large area, dense arrays of nanostructures, and has been used to make arrays of nanoscale magnetic features.[5–7] The BCP pattern can be transferred into the magnetic material using etching, liftoff, a damascene process or other routes.[7]

There have been many examples of the fabrication of BCP-patterned arrays of sub-100 nm diameter dots and anti-dots made from magnetic films with in-plane anisotropy, including Co, Ni and CoFe/Au/NiFe spin valves, as well as films with perpendicular magnetic anisotropy (PMA) made from materials such as CoCrPt, Co/Pd multilayers, $L1_0$-FePt and NdCo, combined with analysis of their collective magnetic and spin transport properties, domain dynamics, reversal mechanism and magnetostatic interactions.[3,4,7–14] However, researchs in magnetic nanowires are much more limited, even though BCP lithography provides a convenient method for fabrication of nanowire arrays. Nanowires with pitch of ~18 nm and above have been made using polystyrene-block-poly(2/4-vinylpyridine) and polystyrene-*block*-polydimethylsiloxane (PS-*b*-PDMS) BCP patterns from metals such as Fe, Co and Ni grown on Si and exhibiting in-plane anisotropy.[5,15,16] PMA CoCrPt nanostructures with 22-44 nm pitch have been patterned using polystyrene-block-polymethylmethacrylate (PS-*b*-PMMA).[17] A smaller pitch can be achieved using BCPs with higher Flory–Huggins interaction parameter such as PS-*b*-PDMS, or by using

PS-*b*-PMMA with self-aligned double patterning.[17] The fabrication and measurement of magnetic nanowire arrays using BCP lithography is less well studied than that of dot arrays, but nanowire arrays hold strong interest for certain devices such as domain wall logic devices[18] and racetrack memory[19] as well as for studies of domain wall behavior at reduced dimensions. In particular, PMA materials such as Co/Pd multilayers and $L1_0$-FePt stand out due to their high magnetocrystalline anisotropy ($K_u$) of $(1.6 - 70) \times 10^6$ erg cm$^{-3}$,[20,21] which promises extremely high areal density of domain walls and high thermal stability of the magnetization,[20–24] but there are no examples of BCP-patterning of line arrays of PMA Co/Pd, $L1_0$-FePt and similar materials.

Closely-spaced arrays of narrow wires with PMA provide model systems for studying not just the scaling of domain walls in wires but the effects of inter-nanowire interactions on the reversal behavior, which is a prerequisite for their applications in high density magnetic devices. There have been studies of the magnetization reversal of nanowire arrays and isolated nanowires of Co, Fe, NiFe and CoFe with in-plane anisotropy,[25–28] and in isolated PMA nanowires such as Co/Pd, Co/Pt and $L1_0$-FePt.[21,29–32] However, studies of PMA nanowires arrays made of CoCrPt (linewidth < 50 nm), Co/Pt (linewidth < 80 nm), and Cu/Ni/Cu (linewidths ≥ 100 nm) generally demonstrated the viability of electron beam, interference or BCP lithography techniques in producing nanowires or analyzed anisotropy contributions to the wires.[17,33–35] Another study of PMA Co/Pt line arrays (linewidths and gap ~200 nm) illustrated Co/Pt magnetic configurations that were used for magnetic force microscopy (MFM) tip calibration.[36] Work on PMA Co/Pd multilayers and $L1_0$-FePt has investigated single nanowires, dot arrays and thin films.[37–40]

In this work, we utilized a BCP patterning technique to fabricate large areas of PMA [Co/Pd]$_{15}$ and $L1_0$-FePt nanowires with 63 nm pitch length and 30-40 nm linewidth. The magnetic hysteresis and the domain nucleation and expansion during reversal were studied by vibrating sample magnetometry, magnetic force microscopy and micromagnetic simulations. Magnetostatic interactions led to correlated domains in adjacent wires with opposite magnetization direction.

**Experimental**

*Co/Pd and $L1_0$-FePt* **Thin** *Film Deposition*:

[Co(0.6 nm)/Pd(1.2 nm)]$_{15}$ was deposited at room temperature on Si substrates with native oxide by magnetron sputtering with Ar pressure of 10 mTorr and a base pressure better than $3 \times 10^{-7}$ Torr. The Co and Pd deposition rates with deposition power of 25 W were 0.62 and 1.86 nm min$^{-1}$, respectively. FePt (20 nm) was ion beam sputtered on a (001) MgO substrate at room temperature from a Fe$_{55}$Pt$_{45}$ target, with Ar pressure of $1.5 \times 10^{-4}$ Torr and base pressure better than $8 \times 10^{-7}$ Torr. The FePt deposition rate was 0.033 nm s$^{-1}$ with an Ar ion gun discharge current, beam current and beam voltage of 0.9 A, 50 mA and 1000 V, respectively. The final composition of the sputtered FePt film was determined to be Fe$_{57}$Pt$_{43}$ from energy dispersive spectroscopy analysis. The PMA $L1_0$-FePt phase was subsequently formed from the as-grown disordered FePt by annealing at 700 °C for 2 hours under a base pressure of $9 \times 10^{-7}$ Torr.

*Block Copolymer Patterning:*

The [Co/Pd]$_{15}$ and $L1_0$-FePt thin films were then patterned using a self-assembled BCP film as am etch mask. A 20 nm thick carbon layer was first electron-beam evaporated on the samples using a power of 0.02 A, voltage of 10 kV, deposition pressure of $3.5 \times 10^{-5}$ mTorr and base pressure $2.2 \times 10^{-5}$ mTorr. The carbon layer provides a hard mask sufficient to protect the magnetic films during Ar ion-beam etching. A PS-*b*-PDMS diblock copolymer with molecular weight of 75.5 kg mol$^{-1}$, polydispersity index of 1.04, and PDMS volume fraction of 0.415 was synthesized and characterized as described in previous work.[41] The PS-*b*-PDMS BCP film (35 nm) was spin-cast on the [Co/Pd]$_{15}$ and $L1_0$-FePt thin films and solvent vapor annealed under flowing nitrogen (2-5 sccm) passing through a glass chamber of volume 88 cm$^3$ containing 4 cm$^3$ of toluene. After two hours of vapor annealing, the film was dried by rapid exposure to air. The BCP films were then reactive-ion etched for 5 s with CF$_4$ to remove the PDMS surface layer (50 W, 15 mTorr, 10 sccm, in a Plasma-Therm 790) followed by 60 s of O$_2$ plasma (90 W, 6 mTorr, 10 sccm) to remove the PS and oxidize the PDMS microdomains. The carbon under the PS was also removed, and the carbon/oxidised-PDMS patterns which were left behind acted as the hard mask. The combination of film thickness and annealing conditions produced a pattern consisting of a layer of cylindrical PDMS microdomains oriented in the plane of the substrate. The cylinders were locally parallel but their in-plane direction varied giving a fingerprint-like pattern.

The patterns were then transferred into the PMA Co/Pd and $L1_0$-FePt thin films by ion milling with an Ar-ion beam source of beam voltage 450 V, acceleration voltage 250 V, Ar pressure of $2 \times 10^{-4}$ Torr and base pressure of $9 \times 10^{-7}$ Torr. The ion milling process for Co/Pd was completed after 180 s when the Co profile from the secondary ion mass endpoint detector (Hiden Analytical) reached a plateau and the two-point probe resistance measurement of the film increased significantly from 120 Ω before etching to 1030 Ω after etching. The $L1_0$-FePt thin film required 200 s etch time before the Fe secondary ion mass profile plateaued and resistance of the film increased from 110 to 580 Ω.

*Structural and Magnetic Analysis:*

The patterned features were imaged by a Zeiss ORION NanoFab helium-ion microscope (HIM). A focused ion beam (FIB, Zeiss Auriga) was used to prepare the samples for scanning transmission electron microscopy (STEM, FEI Titan) cross-sectional microstructure analysis. Magnetic properties of the PMA patterned and unpatterned films were characterized by an ADE model 1660 vibrating sample magnetometer (VSM). Lattice spacing and crystal orientation were studied using X-ray diffraction (XRD) on a Panalytical multipurpose diffractometer with Cu K$_α$ radiation. Feature topography was measured using **Veeco Nanoscope IV** atomic force microscopy (AFM). The magnetic domain structures in the wires were mapped out with magnetic force microscopy (MFM, **Veeco Nanoscope IV**) using a CoCr coated low magnetic moment tip (Bruker MESP-LM). Simulations were performed with the three-dimensional OOMMF micromagnetic solver.[42] Analysis of HIM and MFM images were carried out with the ImageJ processing program.

**Results and discussion**

We first discuss the morphology and microstructure of the BCP-patterned [Co/Pd]$_{15}$ nanowires. **Figure 1(a)** shows an array of parallel cylinders of oxidized PDMS/carbon on a Co/Pd film, prior to metal etching, and Figure 1(c) shows the etched Co/Pd nanowires. Defects such as cylinder terminations and Y-junctions are visible in the BCP pattern.[43] The Co/Pd nanowires had a pitch of 62 ± 3 nm, linewidth of 38 ± 4 nm and a spacing of 26 ± 3 nm. The measured edge root-mean-square roughness ($R_{RMS}$) of the BCP patterns and Co/Pd cylinders after pattern transfer were 1.5 nm and 1.0 nm, respectively, but this measurement is limited by the pixel resolution of the helium-ion microscope (HIM) images and the actual roughness may be lower. Figure 1(e) shows a scanning transmission electron microscopy (STEM) cross-sectional image of the nanowires with a higher resolution inset illustrating a columnar polycrystalline structure. The wires had a width at the top of ~34 nm and spacing ~28 nm similar to the HIM images, but the cross-section reveals a wider footing at the bottom of the wires which extends about 10 nm out of the base of the wires, and a taper angle of 81° ± 4° at the sides of the wires. A slight etch into the Si substrate is also visible. The footings likely consist of redeposited material from the Co/Pd multilayer. Both Co/Pd unpatterned and patterned samples showed a (111) XRD peak (see supplementary information), indicating a preferred texture.

Next, we compare the magnetic properties of the [Co/Pd]$_{15}$ before and after patterning. The magnetic moment per unit area decreased from $8.7 \times 10^{-4}$ emu cm$^{-2}$ (corresponding to saturation magnetization $M_s$ = 320 emu cm$^{-3}$ for the unpatterned film) to $5.8 \times 10^{-4}$ emu cm$^{-2}$ after the removal of Co/Pd magnetic material not covered by the BCP mask. The change in magnetic moment per unit area suggests that 66 % of the Co/Pd thin film was left as nanowires, which is similar to the areal coverage of cylindrical features of 64 % determined from the HIM image [Figure 1(c)]. After patterning, the coercivity ($H_c$), anisotropy ($K_u$) and squareness ($S^*$) decreased significantly from 1990 ± 50 Oe to 790 ± 50 Oe, 1.64 ± 0.20 × 10$^6$ erg cm$^{-3}$ to 0.97 ± 0.14 × 10$^6$ erg cm$^{-3}$ and 0.97 ± 0.23 to 0.43 ± 0.14, respectively [**Figure 2(a) and (b)**]. The $K_u$ is estimated from

$$K_u = \frac{1}{2} M_s H_k$$

, where $M_s$ is the saturation magnetization and $H_k$ is the anisotropy field obtained from the extrapolation of the in-plane magnetization curve to saturation. The switching field distribution (SFD), determined as the full width at half maximum of the first derivative of the partial hysteresis loop ($dM/dH$), also increased from 1700 (± 50) to 2440 (± 50) Oe. The tapering of the Co/Pd film and the distribution of wire sizes and shapes are assumed to lead to a spread in anisotropy and contribute to the SFD.

To clarify the effect of magnetostatic interactions, micromagnetic simulations were performed with the 3-D OOMMF micromagnetic solver to estimate the stray field distribution within a Co/Pd nanowire sandwiched by two neighboring Co/Pd nanowires and its effects on the Co/Pd DW dynamics [**Figure 3(a)-(c)**]. Each Co/Pd nanowire had dimensions of 40 nm wide × 4000 nm long × 27 nm thick, gap between neighboring nanowires of 25 nm, $M_s$ of 323 emu cm$^{-3}$, uniaxial out-of-plane $K_u$ of 0.97 × 10$^6$ erg cm$^{-3}$ and exchange constant of 1.3 × 10$^{-6}$ erg cm$^{-1}$. The nanowire consisted of 6 layers of cells with cell size of 5 nm × 5 nm × 4.5 nm to ensure simulation reliability. The damping constant was set to 0.1 for rapid convergence. The magnetic field at the centre and edge of the center nanowire due to the stray field from its two nearest neighbors magnetized in the same out-of-plane direction was 350 Oe and 516 Oe, respectively [Figure 3(a)].

By comparing the energies of relaxed Néel and Bloch walls in a single Co/Pd nanowire, a Néel wall was taken as the equilibrium structure with a 6.4 % lower energy [Figure 3(a)]. When a domain wall was introduced into the center nanowire, the stray field from the neighboring nanowires promoted DW motion leading to expansion of magnetization anti-parallel to that of the neighboring nanowires, [Figure 3(b,c)], indicating the stray fields will produce antiparallel wire magnetization at remanence. The domains in a specific area of the Co/Pd nanowire array were imaged at remanence after AC demagnetization and after application of different reverse fields [**Figure 4(a)-(e)**]. **Figure 5(a)** shows the magnetic signal superimposed on the nanowire topography for the AFM and MFM data shown in Figure 4(a). Many of the wires showed magnetization direction opposite to their neighbors as expected from the stray field direction. By segmenting the $2 \times 2$ μm$^2$ region in Figure 5(a) into 100 portions of $0.2 \times 0.2$ μm$^2$ each, it was determined that approximately 75 % of the neighbouring Co/Pd nanowires displayed opposite contrast after AC demagnetization. This can be seen in Figure 4(a) where the dark-light contrast in the MFM has double the period of the structure in the AFM, and the locations of DWs in neighboring wires are often adjacent to each other.

This indicates a preferential stable antiparallel magnetization configuration of the Co/Pd nanowires in the AC-demagnetized state as predicted from the micromagnetic simulation. Given the switching field distribution of Co/Pd nanowires (2440 Oe) and that the magnitude of the stray field just from the two closest neighbors (up to 516 Oe) is a large fraction of the $H_c$ of the Co/Pd nanowires (790 Oe), the demagnetizing field is large enough to produce an antiparallel configuration in many of the wires. The correlated domain patterns parallel to the wires are in stark contrast to the domain arrangements in the unpatterned Co/Pd film [Figure 4(f)]. The elongated domains in the Co/Pd nanowires had an average domain length of 460 ± 20 nm while the irregularly shaped domains in the unpatterned film had smaller average domain diameter of 270 ± 20 nm.

Even though most of the nanowires had antiparallel magnetization compared to their nearest neighbors, ~25 % of them displayed a parallel configuration. This is attributed to frustration when the nearest neighbors promote opposite magnetization of the center wire, giving degenerate ground states susceptible to small perturbations.[44,45]

To illustrate the reversal process of the array from saturation, the nanowires were first saturated with an out-of-plane field of $H_z$ = 12 kOe giving a magnetization 'down' (yellow or light contrast) configuration. At a reverse field of $H_z$ = -500 Oe [Figure 4(b)], reversed 'up' (red or dark contrast) domains with average size of 120 ± 20 nm nucleated in some of the nanowires. Specific reverse domains are identified (circled) so that changes in their geometry with increasing field can be seen. At an applied field of $H_z$ = -1000 Oe [Figure 4(c)], the existing 'up' domains, with average domain size of 190 ± 20 nm, expanded slightly but the clearest effect is an increase in the density of 'up' domains. These observations suggest a limited domain wall mobility in the film such that domain nucleation occurs at lower fields than are required to translate the walls.

At $H_z$ = -1500 Oe [Figure 4(d)], the reverse domains expanded along the wires, reaching an average size of 540 ± 20 nm. Many regions of adjacent wires (e.g. circled on the right of the image) showed antiparallel magnetization between neighbors as was seen in the AC-demagnetized sample. The stray fields combined with the applied field encouraged expansion of the 'up' domains along the Co/Pd nanowires as predicted in the micromagnetic model to produce the antiparallel magnetization state. At applied fields above $H_z$ = -2000 Oe [Figure 4(e)], the 'up' domains further propagated leaving isolated 'down' domains stabilized by the stray fields.

A related behavior was seen in the unpatterned film where reverse domains formed at -500 Oe and further domains nucleated at -1500 Oe with little growth of the existing domains [Figure 4(g,h)]. Larger field led to expansion of the reverse domains [Figure 4(i)-(j)] and at -4000 Oe only a few 'down' domains remained. The unpatterned and patterned films can be compared by plotting the fraction of reverse domains at remanence vs. field normalized to the saturation field, taken as 4.8 kOe for the unpatterned film and 3.2 kOe for the patterned film [Figure 5(c)]. For the initial stages of reversal the patterned film had significantly more area fraction of reversed domains suggesting that that domain nucleation was relatively easier in the nanowires, presumably facilitated at the edges of the wires, but for larger fields the reverse domains expanded proportionately more in the unpatterned film as they were unconstrained by the wire edges.

We now describe the behavior of nanowire array made from a $L1_0$-FePt film. The BCP film on the $L1_0$-FePt thin film formed a less regular array of microdomains consisting of a mixture of cylinders, perforated lamellae [Figure 1(b)] and partially interconnected double layer lamellae [Figure 1(b) inset], compared to the BCP film on Co/Pd deposited and annealed under the same conditions. The BCP patterns on the $L1_0$-FePt had poor long-range ordering and larger rms line-edge roughness of 3.7 nm. The difference is believed to originate from variations in thickness of the BCP film on the $L1_0$-

FePt thin film, which dramatically affects the BCP morphology.[42] The $L1_0$-FePt thin film had a granular surface with high rms roughness of 2.9 (± 0.5) nm (for comparison, the Co/Pd roughness was ~0.8 ± 0.1 nm). In the regions of the $L1_0$-FePt film where nanowires had formed, their linewidth was 27 ± 8 nm, nanowire spacing was 30 ± 9 nm, centre-to-centre period was 63 ± 8 nm and measured rms edge roughness was 3.4 nm [Figure 1(d)].

The unpatterned $L1_0$-FePt thin film had an out-of-plane easy axis with $H_{c,OP}$ of 1.86 ± 0.05 kOe and an in-plane hysteresis loop with $H_{c,IP}$ of 1.12 ± 0.05 kOe [Figure 2(c)]. The XRD spectra of the unpatterned FePt thin film shows the FePt (001) superlattice peak and (002) fundamental peak (in supplementary information), originating from ordering of the alternating Fe and Pt planes giving a (001)-textured fct $L1_0$-FePt. The extent of ordering is represented by the chemical ordering parameter,

$$S \propto \left(\frac{I_{001}}{I_{002}}\right)^{\frac{1}{2}}$$

, where $I_{001}$ and $I_{002}$ are the integrated intensities of the (001) and (002) peaks. The unpatterned $L1_0$-FePt film gave $I_{001}/I_{002} = 0.45$ indicative of incomplete ordering of the Fe and Pt which is consistent with the in-plane hysteresis. The unpatterned film had $M_s$ = 644 ± 30 emu cm$^{-3}$, $S^*$ = 0.50 ± 0.14, $K_u$ = 6.47 ± 0.32 × 10$^6$ erg cm$^{-3}$ and SFD = 7.81 ± 0.05 kOe.

Upon BCP pattern transfer onto the $L1_0$-FePt thin film, the magnetic moment per unit area decreased from 1.29 to 0.71 × 10$^{-3}$ emu cm$^{-2}$. This suggests that 55 % of $L1_0$-FePt is left after patterning, which exceeds the estimated 46 % areal coverage of cylindrical features calculated from the HIM image [Figure 1(d)]. This can be explained by morphological variations in the BCP, in particular regions of double layer cylinders which led to unpatterned areas of the film. There was a split of the (002) fundamental peak into two distinct (200) and (002) peaks after patterning [Figure 2(c) inset]. This suggests the presence of disordered A1-FePt phase formed from the sidewall redeposition of ion-milled materials during pattern transfer, which could also have contributed to the reduction of the nanowire anisotropy $K_u$ to 2.92 ± 0.26 × $10^6$ erg $cm^{-3}$ [Figure 2(d)]. The $H_{c,OP}$, $H_{c,IP}$ and $S^*$ of the patterned nanowires also increased to 3.79 ± 0.05 kOe, 4.34 ± 0.05 kOe and 0.88 ± 0.18, respectively, presumably affected by the relatively large edge roughness, the inhomogeneity in the nanowire morphology and the crystal structure.

To examine the influence of dipolar stray field on the reversal process, OOMMF simulations were carried out on FePt nanowires with a dimension of 30 nm wide × 4000 nm long × 20 nm thick, a gap between adjacent nanowires of 30 nm, $M_s$ of 644 emu $cm^{-3}$, uniaxial out-of-plane $K_u$ of 2.92 × $10^6$ erg $cm^{-3}$, exchange constant of 1.3 × $10^{-6}$ erg $cm^{-1}$ and damping constant of 0.1 [Figure 3(d)-(f)]. Each nanowire consisted of 4 layers of cells with cell size of 5 nm × 5 nm × 5 nm. The total dipolar stray field experienced by the center $L1_0$-FePt nanowire along its centre and edge was 445 and 583 Oe, respectively [Figure 3(d)]. Different from Co/Pd simulation results, both Néel and Bloch walls initiated in the $L1_0$-FePt nanowire relaxed to give a Bloch wall [Figure 3(d)], which was taken as the energetically favourable DW structure for subsequent simulation [Figure 3(e) and (f)]. In the absence of an external field and omission of pinning from defects and edge roughness, the stray field emitted from the 'up' neighbouring nanowires encouraged the expansion of the spin 'down' domain and propagation of the Bloch wall to the left [Figure 3(f)].

Similar experimental protocols were used to examine the reversal of the $L1_0$-FePt nanowires as were used for the Co/Pd. A superimposed AFM-MFM image after AC-demagnetization [Figure 5(b)] shows that an estimated 40% of the nearest-neighbor $L1_0$-FePt nanowires were antiparallel to their neighbors, less than observed for the Co/Pd sample. OOMMF calculations indicated that dipolar stray field (445 - 583 Oe) from the neighbouring $L1_0$-FePt nanowires was much smaller than the coercivity $H_{c,OP}$ (3.79 kOe) and hence the stray fields are less effective in determining the magnetization state. The AC-demagnetized $L1_0$-FePt nanowire sample had an average domain size of 425 ± 20 nm [**Figure 6(a)**] which was similar to that of the uncorrelated labyrinth domain configurations in the unpatterned $L1_0$-FePt thin film with average domain size of 450 ± 20 nm [Figure 6(f)].

After saturation at $H_z$ = 12 kOe, giving an initial 'down' configuration (yellow contrast), an applied field of $H_z$ = -2 kOe [Figure 6(b)] led to nucleation of reverse 'up' domains (red) which expanded at larger reverse fields [Figure 6(c)-(e)]. There was no clear correlation of domains between adjacent wires, unlike in the Co/Pd sample. The reversal behaviour was similar to that of the unpatterned $L1_0$-FePt thin film [Figure 6(g) and supplementary information]. Furthermore, the area fraction of reversed domains vs. field was similar for the patterned and unpatterned $L1_0$-FePt film [Figure 5(d)], suggesting that the reversal is dominated by intrinsic domain nucleation and growth rather than by the pattern geometry.

**Conclusions**

In conclusion, PMA [Co/Pd]$_{15}$ and $L1_0$-FePt nanowire arrays with period 63 nm, linewidths of 27 – 38 nm, thickness of 20 – 27 nm and wire spacings of 26 – 30 nm were fabricated by ion beam etching using a carbon hard-mask patterned by a self-assembled PS-*b*-PDMS diblock copolymer mask, and the magnetic properties and reversal process were characterized. This process produced wires with a high aspect ratio (thickness/wire width = 0.71 – 0.74) with a predicted Néel domain wall structure in the Co/Pd and a Bloch wall in the FePt nanowires. In both cases dipolar stray fields from nearest neighbor wires are predicted to drive domain wall motion producing antiparallel magnetization directions in adjacent nanowires, but the ratio of stray field to coercivity is much higher for Co/Pd compared to FePt. Co/Pd nanowires showed a highly correlated domain structure in which adjacent wires had antiparallel magnetization and domain wall locations were aligned, as a result of the magnetostatic interactions. The study revealed domain nucleation was the dominant process at lower reverse fields with domain wall propagation occurring at higher reverse fields, i.e. the nanowires exhibit limited domain wall mobility. The $L1_0$-FePt nanowires were less regular due to the effect of the higher film roughness on the BCP morphology, and the stray field was much smaller than the coercivity. This led to only a limited correlation between the magnetization directions of nearest neighbor nanowires.

**Acknowledgements**

The authors gratefully acknowledge the support of the Agency of Science, Technology and Research (A*STAR) International Fellowship grant, and C-SPIN, a pattern transfer. Inset of (c) shows a cross-sectional HIM image of the tapered Co/Pd nanowires. (d) $L1_0$-FePt nanowires and other structures after pattern transfer. (e) Cross-sectional TEM image of BCP patterned Co/Pd nanowires. Inset shows the polycrystalline lattice of a Co/Pd nanowire, where the grain boundary is depicted by white symbols.

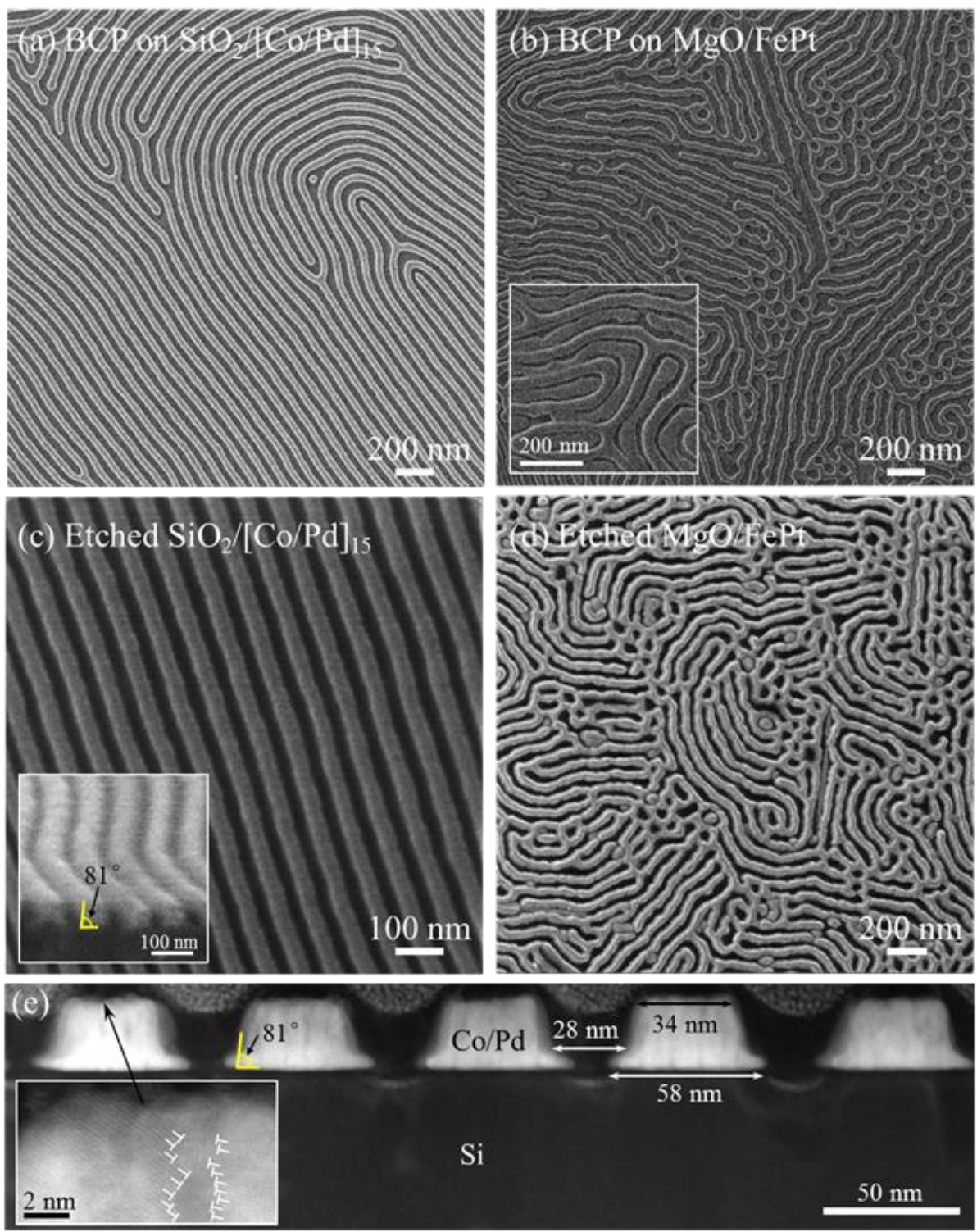

Figure 1. Plan-view HIM images showing (a) BCP patterns on $SiO_2/[Co/Pd]_{15}$ and (b) BCP patterns on MgO/$L1_0$-FePt. Inset of (b) shows a double layer BCP lamellar structure. (c) Co/Pd nanowires after

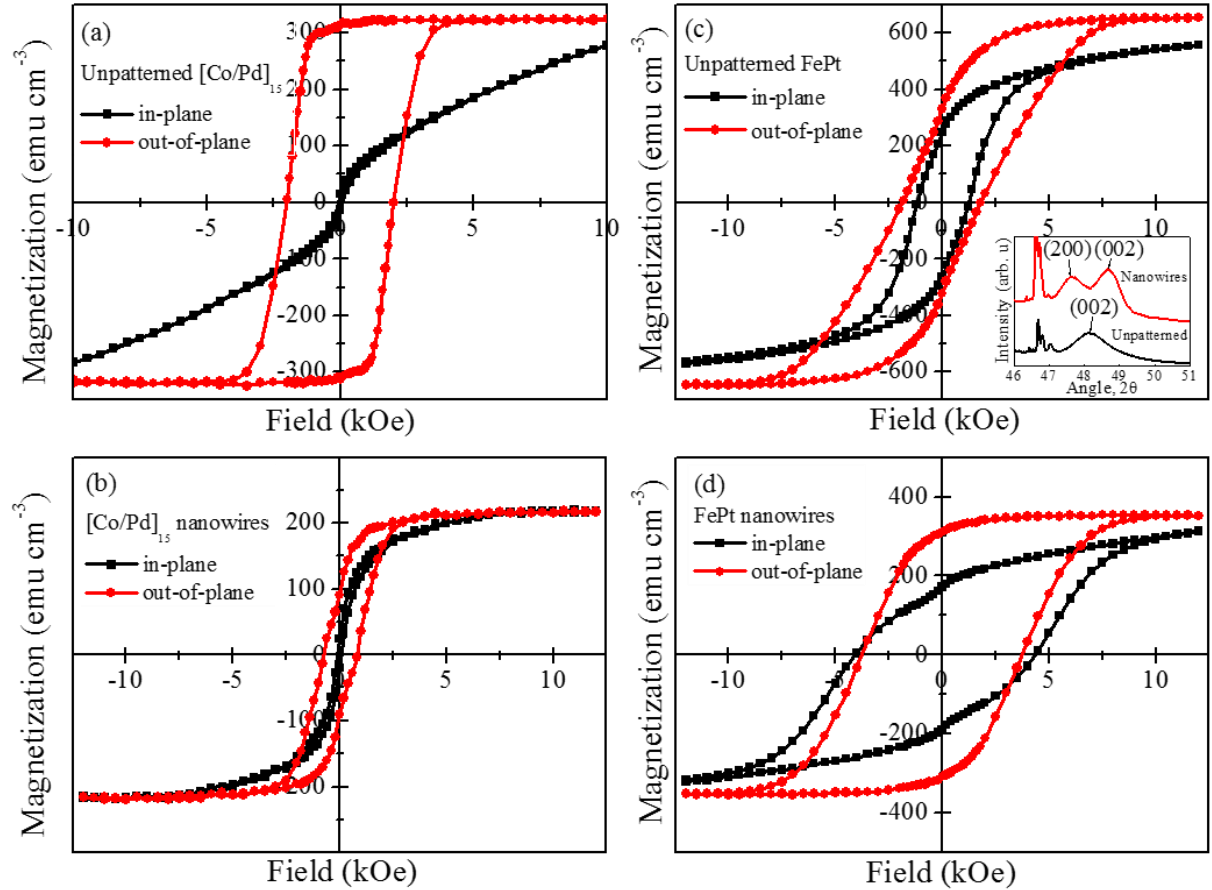

Figure 2. Hysteresis loop of (a) unpatterned SiO$_2$/[Co/Pd]$_{15}$ thin film, (b) BCP patterned array of Co/Pd nanowires, (c) unpatterned MgO/$L1_0$-FePt (20 nm) thin film. Inset shows the XRD spectrum of the $L1_0$-FePt before and after BCP patterning. The unlabelled sharp peaks originate from the MgO substrate. (d) BCP patterned array of $L1_0$-FePt nanowires.

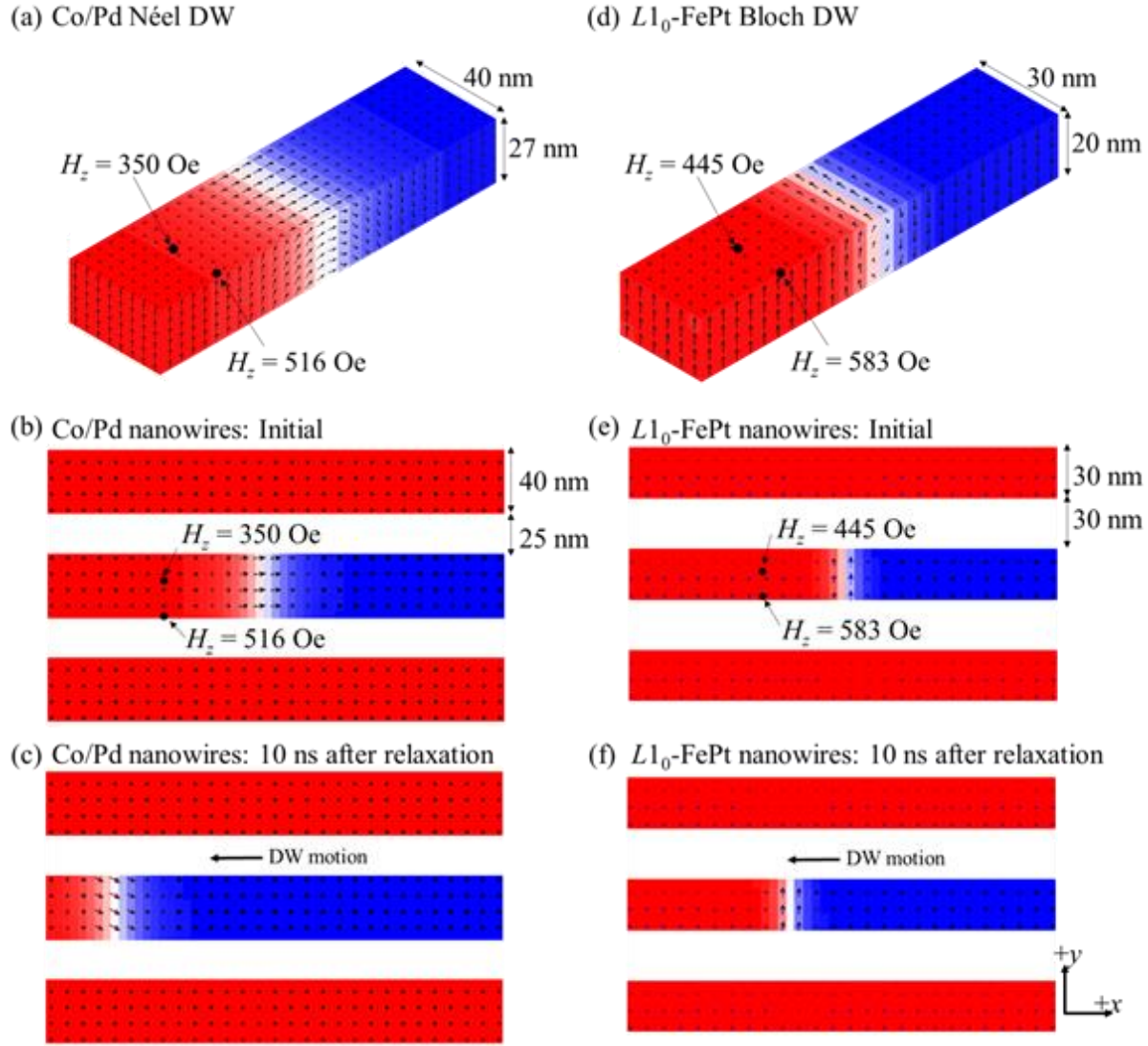

Figure 3. OOMMF simulations. In each case red and blue represent the component of magnetization out of the film plane and the arrows show magnetization vectors. (a) 3D schematic of a Co/Pd Néel DW and the magnitude of dipolar stray field experienced at the edge and centre of the Co/Pd nanowire from its two nearest neighbors. (b) Plan-view of three Co/Pd nanowires, with a Néel DW in the center nanowire, and the outer nanowires magnetized 'up' (red). (c) Propagation of the 'down' domain (blue) by Néel DW motion to the left in the center Co/Pd nanowire driven by the stray field from its neighbors. (d) 3D schematic of a $L1_0$-FePt Bloch DW and the magnitude of dipolar stray field experienced at the edge and centre of the $L1_0$-FePt nanowire from its two nearest neighbors. (e) Plan-view of three $L1_0$-FePt

nanowires, with a Bloch DW in the center nanowire, and the outer nanowires magnetized 'up' (red). (f) Propagation of the 'down' domain (blue) by Bloch DW motion to the left in the center $L1_0$-FePt nanowire driven by the stray field from its neighbors.

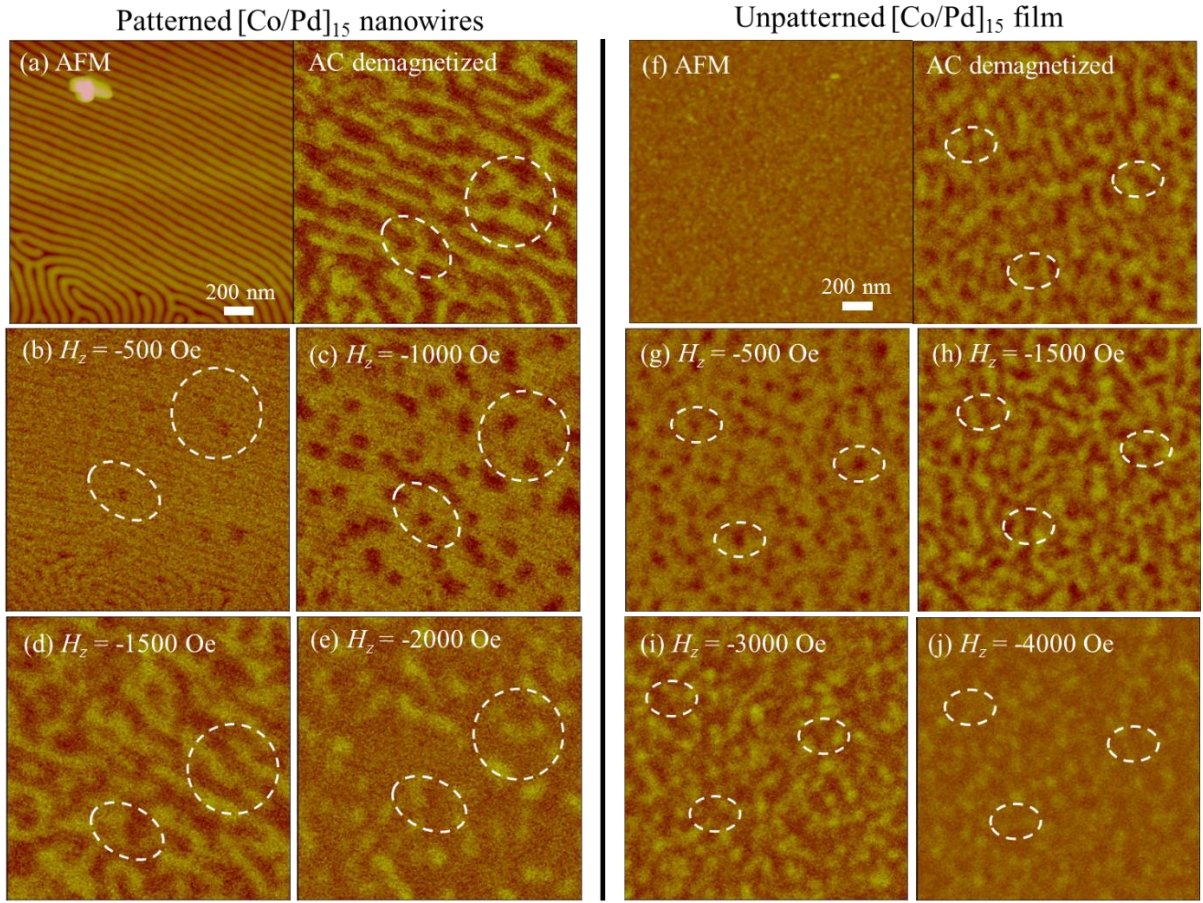

Figure 4. (a) AFM (left) and MFM (right) images of BCP patterned Co/Pd nanowires after AC demagnetization. (b-e) MFM images of Co/Pd nanowires at remanence after applying a field of $H_z$ = (b) -500, (c) -1000, (d) -1500 and (e) -2000 Oe. (f) AFM (left) and MFM (right) images of unpatterned Co/Pd thin film after ac-demagnetization. (g-j) MFM images of unpatterned Co/Pd thin film at remanence after applying a field of $H_z$ = (g) -500, (h) -1500, (i) -3000 and (j) -4000 Oe. All MFM images are captured at a fixed location on the patterned and unpatterned Co/Pd. Red and yellow contrast in the MFM images indicate 'up' and 'down' stray fields, respectively.

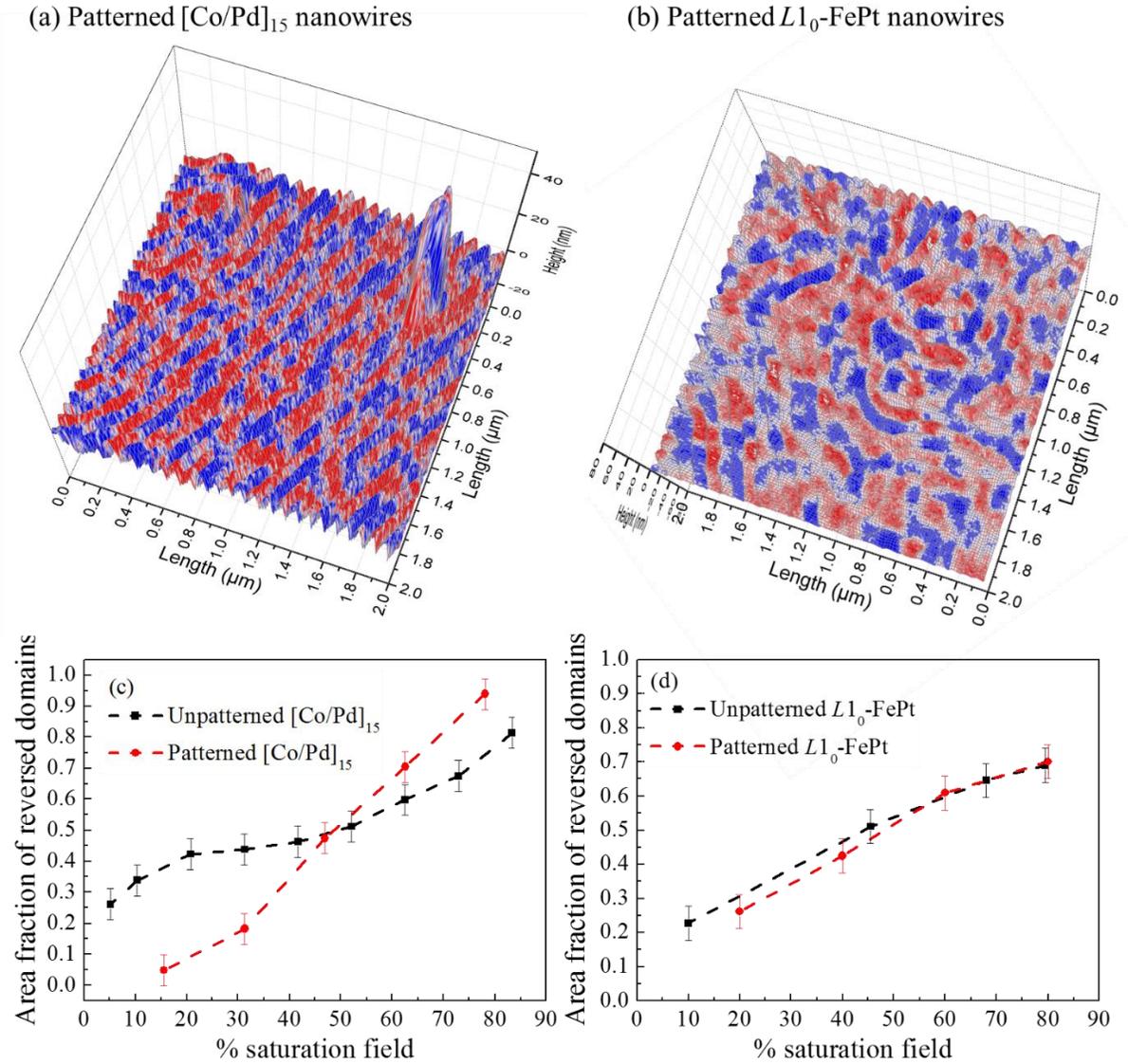

Figure 5. (a,b) 3D diagram showing the magnetic information (MFM) superimposed on the topography (AFM) of the ac-demagnetized (a) Co/Pd and (b) $L1_0$-FePt nanowires. Blue and red shading indicate 'up' and 'down' magnetization, respectively. (c,d) Area fraction of reversed domains versus applied reversed field as a percentage of saturation field for unpatterned and BCP patterned films of (c) Co/Pd and (d) $L1_0$-FePt. The dashed lines serve as a guide for the eye. The saturation field is taken to be the field at which the out-of-plane hysteresis loop closes. Saturation fields of the unpatterned Co/Pd, Co/Pd nanowires, unpatterned FePt film and FePt nanowires are 4.8, 3.2, 8.8, and 10.0 kOe, respectively.

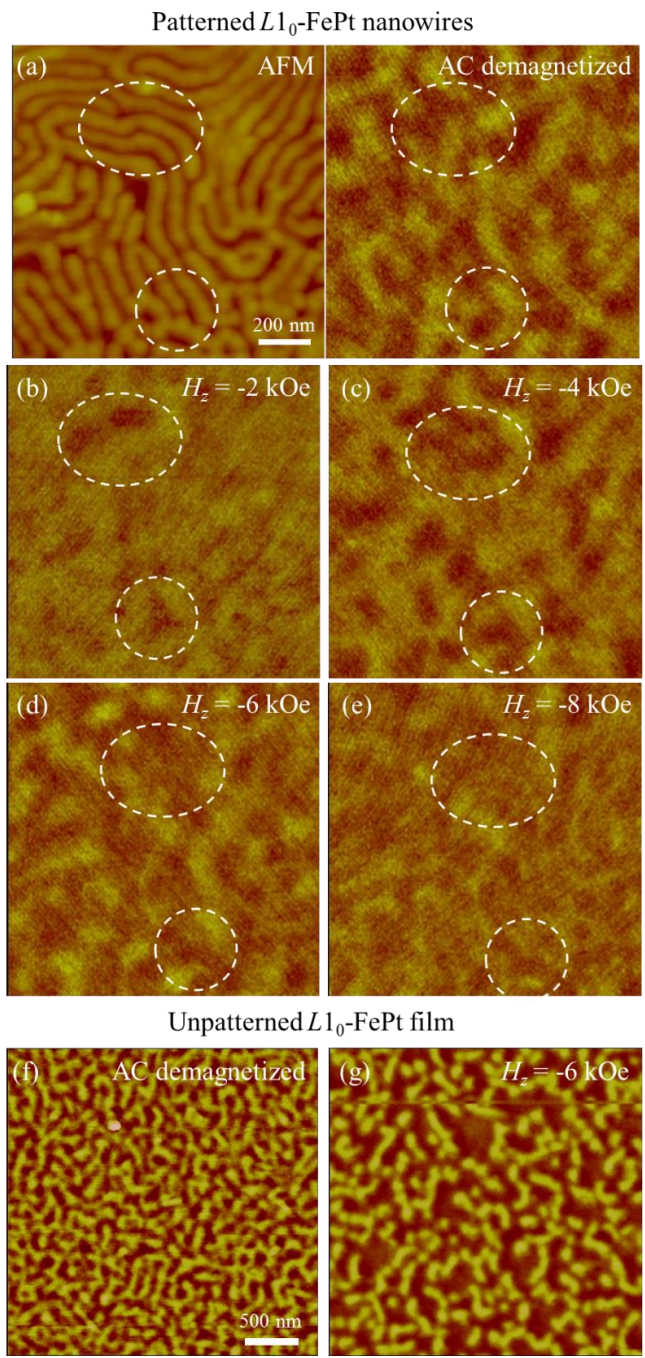

Figure 6. (a) AFM (left) and MFM (right) images of BCP patterned $L1_0$-FePt nanowires after ac-demagnetization. MFM images with applied field of $H_z =$ (b) -2, (c) -4, (d) -6, (e) -8 kOe, captured at a fixed location on the sample. MFM images of unpatterned $L1_0$-FePt after (f) ac-demagnetization and (g) applied field of $H_z =$ -6 kOe. Red and yellow contrast in the MFM images indicate 'up' and 'down' magnetization, respectively.